\documentstyle[12pt,epsfig]{article}

\newcommand{\ppp}[1]{%
        \setbox0=\hbox{#1}%
        \kern-.02em\copy0\kern-\wd0
        \kern+.04em\copy0\kern-\wd0
        \kern-.02em\raise.0217em\box0}
\newcommand{\vek}[1]{
        \mathchoice{\mbox{\boldmath$#1$}}%
        {\mbox{\boldmath$#1$}}%
        {\ppp{$\scriptstyle#1$}}%
        {\ppp{$\scriptscriptstyle#1$}}}

\newcommand{\lsim}{$\raisebox{-0.8ex} {$\stackrel{\textstyle <}{\sim}$}$}

\begin{document} 


\title{\Large\bf Resonances and higher twist in polarized lepton-nucleon
scattering\footnote{Work supported in part by DFG and BMBF}
}

\bigskip

\author{J.~Edelmann, G.~Piller, N.~Kaiser and  W.~Weise}
\date{\today{}}

\maketitle

\begin{center}
\vspace*{1cm}

Physik-Department, Technische Universit\"{a}t M\"{u}nchen, \\
D-85747 Garching, Germany 
\end{center}

\vspace*{5cm}

\begin{abstract} 
We present a detailed analysis of resonance contributions in the context 
of higher twist effects in the moments of the proton spin structure function 
$g_1$. For each of these moments, it is found that there exists a
characteristic $Q^2$ region in which (perturbative) higher twist corrections 
coexist with (non-perturbative) resonance contribution of comparable
magnitude.
\noindent
\bigskip


\end{abstract}
\thispagestyle{empty}
\newpage

\section{Introduction}
High energy lepton scattering is a well established tool to investigate the
structure 
of the nucleon. We restrict ourselves to charged leptons ($e$ or $\mu$); 
the exchanged virtual photon transfers four-momentum $q^{\mu}=(q_0,\vek{q})$,
with the resolution 
determined by the virtuality $Q^2=-q^2=\vek{q}^2-q_0^2$. At
$Q^2\gg 1 {\rm GeV}^2$ deep-inelastic scattering resolves the partonic 
constituents (quarks and gluons) of the nucleon. At $Q^2\;\lsim\; 1
{\rm GeV}^2$, on the other hand, the excitation of nucleon resonances and
multi-pion continuum states is important. Exploring the transition between
 partonic and hadronic scales is of great significance to our
understanding of the nucleon.
The aim of the present paper is to discuss 
polarized 
lepton-nucleon scattering in kinematic regions where both hadron and parton
degrees of freedom are expected to coexist. 

The response of the nucleon is
expressed in terms of the hadronic tensor
\begin{eqnarray} \label{eq:Wmunu}
W_{\mu\nu}(x,Q^2)&=&\frac{1}{4\pi}\sum_X
(2\pi)^4\delta^4(P\!+\!q\!-\!P_X)
\langle N(P,S)|J_{\mu}(0)|X(P_X,\lambda_X)\rangle\nonumber
\\&&
\langle X(P_X,\lambda_X)|J_{\nu}(0)|N(P,S)\rangle\,\\
&=& 
\label{eq:strfns}
W_{\mu\nu}^{(S)}+W_{\mu\nu}^{(A)}\nonumber.
\end{eqnarray}

The matrix elements of the electromagnetic current $J_{\mu}$
describe the transition of a nucleon with four-momentum $P$, invariant mass $M$
($P^2=M^2$) and spin $S$ to a hadronic final state $X$ with four-momentum
$P_X$ and polarization $\lambda_X$. The sum in (\ref{eq:Wmunu}) implies an
integration over three-momentum, $\frac{d^3P_X}{(2 \pi)^32P_{X0}}$, and the normalization of
$|N\rangle$ and $|X\rangle$ is $\langle N(P',S)|N(P,S)\rangle=2P_0(2\pi)^3 \delta^3(\vek{P}'-\vek{P})\delta_{S,S'}$.

The symmetric part $W_{\mu\nu}^{(S)}$ involves the spin independent structure
functions $F_{1,2}$  measured in the scattering of unpolarized particles. The
antisymmetric term,
\begin{eqnarray}
W_{\mu\nu}^{(A)}=
i \epsilon_{\mu\nu\lambda\sigma} q^{\lambda} 
\left[
\frac{g_1(x,Q^2)}{P \cdot q} S^{\sigma}
+\frac{g_2(x,Q^2)}{(P \cdot q)^2}
(P \cdot q\; S^{\sigma}-q\cdot S\; P^{\sigma}) \right]. 
\end{eqnarray}
introduces the spin structure functions $g_1$ and $g_2$.
The nucleon spin vector is
$S^{\sigma}=\frac{1}{2}\bar{u}(P,S)\gamma^{\sigma}\gamma_5 u(P,S)$
with Dirac spinors normalized as $\bar{u}u=2M$. The structure functions depend
on the Bjorken variable $x=Q^2/(2 P\cdot q)$ and on $Q^2$.

Spin structure function data have been taken at SLAC, CERN
and DESY 
\cite{E143,E154,SMC,Hermes}, primarily in the partonic high $Q^2$ range.
Polarized deep-inelastic scattering in the resonance region was measured by 
the E143 collaboration at SLAC \cite{E143}.
In the first part of our study we combine these data with other available
information from the photo- and leptoproduction of nucleon resonances and
investigate their contribution to the moments of the proton spin structure
function $g_1$.

The influence of resonances and non-resonant low-mass excitations turns out to
be quite significant for $Q^2\;\lsim\; 4\;{\rm GeV}^2$, as we shall demonstrate.
For example, at $Q^2\;=\;2\;{\rm GeV}^2$ they account for as much as $20\%$ of the
first moment of $g_1$. Similar observations have been made for unpolarized
deep-inelastic scattering \cite{JiUnrauUnpol}.

In the second part we use the QCD operator
product expansion and extract twist-4 matrix elements from the leading moments
of $g_1$. For the first moment such an analysis has been carried out in great
detail in   ref.\cite{JiMel}. 
We find substantial higher-twist contributions to the
first, third and fifth moments of $g_1$ for $Q^2\;\lsim \;2,4 $ and $10\;{\rm
GeV}^2$, respectively. We examine target mass effects and
investigate the different components of the higher-twist pieces of $g_1$. 
It turns out that contributions from elastic scattering, low-mass hadronic
excitations and the partonic high-mass continuum are all of similar 
importance. We comment on the applicability of the twist expansion and
recall basic ideas of parton-hadron duality. Altogether our results emphasize
the need for high-precision experiments in the resonance region, to be
performed at the Jefferson laboratory \cite{JEFPROP}. 

\section{Twist expansion of \large
$\lowercase{{g_1}}$}

In this section we briefly summarize results from  the operator product expansion 
for the nucleon spin structure function $g_1$ 
(for details see e.g. \cite{Shuryak Vains81}).  
Following the conventions of  ref.\cite{Ehren94} we introduce the 
$n$-th moment of $g_1$ as:
\begin{equation} \label{eq:g1_moments}
g_{1}^{(n)} (Q^2) =\int_0^1{\rm d}x \,x^{n-1} g_{1}(x,Q^2)
\quad\quad (\rm{with}\;\;\; n=1,3,5\dots).
\end{equation}
Note that the upper limit of integration includes the contribution 
from elastic scattering. 
Its presence  results from the fact that the operator product 
expansion, applied to deep-inelastic scattering, 
implicitly involves a sum over all final hadronic states including the 
nucleon itself. 
The importance of the elastic component  in a QCD analysis of 
structure function moments has been emphasized in  ref.\cite{JiUnrauUnpol}.

At large momentum transfers, $Q^2 \gg \Lambda_{\rm{QCD}}^2$,  
the moments (\ref{eq:g1_moments}) can be written in terms of the twist expansion
\cite{JiUnrauG1}: 
\begin{equation}\label{eq:twist_expansion}
g_{1}^{(n)} (Q^2) = \sum_{\tau = 2,4,\dots} 
\frac{\mu_{\tau}^{(n)}(Q^2)}{Q^{\tau - 2}}.
\end{equation}
"Twist" is a useful bookkeeping device to classify the light
cone singularity of the coefficients in the QCD operator product expansion.
Let a local operator in this expansion be a Lorentz tensor of rank $r$
with (mass) dimension $d$, and let $\sigma\le r$ be the "spin"associated with
this operator. Then twist is defined as $\tau=d-\sigma$.
The functions $\mu_{\tau}^{(n)}(Q^2)$ 
are related to  nucleon matrix elements of quark and gluon operators 
with maximal twist $\tau$.  
Their leading (logarithmic) $Q^2$-dependence can be calculated perturbatively 
as a series expansion in the strong coupling constant 
$\alpha_s$.
It should be mentioned that, due to the asymptotic nature of 
QCD perturbation theory, a systematic separation of the twist expansion 
and the perturbation series for 
$\mu_{\tau}^{(n)}$ is non-trivial and 
 still a matter of ongoing investigations  
(for detailed discussions see e.g  ref.\cite{Mueller:1993pa}). 

Up to corrections of order $1/Q^{4}$ one finds  
:
\begin{eqnarray} \label{eq:g1n_HT}
g_1^{(n)}(Q^2)&=&
\frac{1}{2} a_{n-1}(Q^2)+\frac{M^2}{Q^2}\frac{n(n+1)}{2(n+2)^2}
\left(n\, a_{n+1}(Q^2)+4 \,d_{n+1}(Q^2)\right)\\ &+& 
\frac{4}{9}\frac{M^2}{Q^2} f_{n+1} (Q^2)
+{\cal O}\left(\frac{M^4}{Q^4}\right)
\nonumber
\end{eqnarray}
The coefficients $a_n$ represent the genuine twist-$2$ contributions to  
$g_1^{(n)}$. They  depend only logarithmically on $Q^2$ and  
dominate for $Q^2$ much larger than a typical hadronic scale, 
say the squared nucleon mass  $M^2$. 
The second term in (\ref{eq:g1n_HT}) arises from  target 
mass corrections \cite{PiccioneRidolfi}. 
They are determined  by the twist-$2$ pieces $a_n$ and the twist-$3$ 
corrections $d_n$ related to moments of the 
spin structure function $g_2$
\cite{Jaffe:1991qh}
:
\begin{equation} \label{eq:dn_g2}
d_{n-1} = 2\,g_1^{(n)} + \frac{2 n}{n-1} \, g_2^{(n)} + 
{\cal O}
\left(\frac{M^4}{Q^4}\right).
\end{equation}
The true twist-$4$  contributions in eq.(\ref{eq:g1n_HT}) are denoted by $f_{n+1}$.
For higher moments, $n>1$, several matrix elements of 
twist-$4$ are involved. 
Their sum gives the coefficient $f_{n+1}$ in (\ref{eq:g1n_HT}).

In our work twist-$2$ contributions  
are defined through  moments of 
presently available NLO  
parametrizations  of $g_1$ \cite{Gehrmann}.
The extraction of higher twist contributions from structure function data 
has been a subject of  recent studies \cite{JiMel}. 
The active interest in these quantities derives from the fact that they  are 
related to matrix elements which are sensitive to  
quark-gluon interactions in the nucleon. 
For example, one has \cite{Shuryak Vains81,Ji:1995bs}: 
\begin{equation}
2 f_2(Q^2) M^2 S^{\mu} = 
\sum_f e_f^2 \langle N(P,S) | g \bar \psi_f \tilde G^{\mu\nu} \gamma_{\nu} 
\psi_f |N(P,S) \rangle.
\end{equation} 
The sum is taken over all quark fields $\psi_f$ with 
flavor $f$  and charge $e_f$, and
$\tilde G^{\mu\nu}$ stands for the dual gluon field 
strength tensor ($g$ denotes the QCD coupling strength).


\section{Helicity amplitudes}

In this paper we investigate contributions to the 
proton spin structure 
function $g_1^p$  resulting from the electro-production of nucleon  
resonances, as well as from the production of continuum states 
in the deep-inelastic regime.
Resonance contributions are conveniently described in terms of
helicity amplitudes \cite{Carlson:1998ji}:
\begin{equation} \label{eq:Gm}
G_m=\frac{1}{2M}
\left\langle X(P_X, \lambda'=m-\frac{1}{2})\right|
\epsilon^{m}\cdot J(0)
\left|N(P,\lambda=-\frac{1}{2})\right\rangle.
\end{equation}
We choose $\vek{q}/|\vek{q}|$ as the spin quantization axis.
The amplitude $G_m$ represents the production of a hadronic state $X$
with spin projection $\lambda'$ following the absorption of a virtual 
photon with polarization (helicity) $m=\pm 1,0$ on a nucleon with spin projection
$\lambda = -1/2$. 
The photon polarization vectors are 
$\epsilon^{\pm}=(0,\mp1,-i,0)/\sqrt{2}$ and 
$\epsilon^{0}=(|\vek{q}|,0,0,\nu)/{Q}$, with $Q = \sqrt{Q^2}$.

Combining eqs.(\ref{eq:Wmunu}) and (\ref{eq:Gm}) gives:
%
\begin{eqnarray}\label{statex}
g_{1}=\frac{1}{1+\frac{Q^2}{\nu^2}}\sum_{X} M^2\delta(W^2-M_X^2)
\left[ |G_+|^2-|G_-|^2+\frac{\sqrt{2 \,Q^2}} {\nu} 
\,G_0^* \,G_+\right],
\end{eqnarray}
with $\nu=P\cdot q/M$.
The final state $X$ with invariant mass $M_X$ has 
$\lambda'=+1/2$ for $G_{+}$, $\lambda'=-3/2$ for $G_{-}$, and  $\lambda'=-1/2$ for $G_0$. 
It is common to use the amplitudes ($e$ is the electric charge with $e^2/4\pi = 1/137$):
\begin{equation}\label{eq:A's}
A_{1/2}=e\,\sqrt{\frac{M}{W^2-M^2}}\,G_+ ,\;\;
A_{3/2}=e\,\sqrt{\frac{M}{W^2-M^2}}\,G_- ,\;\;
S_{1/2}=e\,\sqrt{\frac{M}{W^2-M^2}}\frac{|\vek{q^*}|}{Q}\,G_0, 
\end{equation}
where $\vek q^*$ denotes the three-momentum transfer as measured in the 
photon nucleon center-of-mass frame, 
i.e. ${\vek q^*}^2=Q^2+(W^2-M^2-Q^2)^2/4 W^2$ with the
total c. m. energy $W$.

\section{Model}
\label{sed:Model}

In the following we present a parametrization of  the proton 
structure function $g_1^p$ which is applicable at 
small and moderate values of $Q^2$. 
We follow hereby closely an analysis of recent  
data in  ref.\cite{E143}.  
At small center-of-mass energies, $W < 1.7$ GeV,  we account for 
the contribution of dominant nucleon resonances. 
In addition, a phenomenological non-resonant background is added. 
For large  $W > 1.7$ GeV we use  an existing parametrization 
of available  data.

The contribution of an isolated nucleon resonance to  
$g_1$ is usually  expressed through
helicity dependent virtual photon-nucleon cross sections. 
In terms of the helicity amplitudes  (\ref{eq:A's}) these are 
defined as:
\begin{eqnarray} \label{eq:photon-nucleon_cross_sections}
&&\sigma^{T}_{1/2,3/2}=\frac{M\Gamma_{R} }{M_{R}[(W-M_{R})^2+\Gamma_{R}^2/4]}
 |A_{1/2,3/2}|^2,\nonumber \\
&&\sigma^{L}_{1/2}=\frac{M\Gamma_{R} }{M_{R}[(W-M_{R})^2+\Gamma_{R}^2/4]}\frac{Q^2}{\vek{q^*}^2}|S_{1/2}|^2,\nonumber \\
&&\sigma^{LT}_{1/2}=
\frac{M\Gamma_{R} }{\sqrt{2} M_{R}[(W-M_{R})^2+\Gamma_{R}^2/4]}
\frac{Q}{|\vek{q^*}|}
\,S_{1/2}^*A_{1/2}.
\end{eqnarray}
Here $M_{R}$ is the mass and $\Gamma_{R}$ the width of the resonance. 
Combining eqs.(\ref{statex},\ref{eq:A's},%
\ref{eq:photon-nucleon_cross_sections}) gives for the contribution 
of a resonance $R$ to $g_1$:
\begin{eqnarray}\label{g1res}
\left.g_1(x,Q^2)\right|_{R} &=&
\frac{\nu M-Q^2/2}{4 \pi^2 \alpha} \frac{1}{1+Q^2/\nu^2}
  \left(\frac{\sigma^{T}_{1/2}-\sigma^{T}_{3/2}}{2}
+\frac{Q}{\nu}\sigma^{LT}_{1/2}\right), 
\end{eqnarray} 
where the photon-nucleon cross sections refer to the excitation 
of $R$. 
At low $W$ the helicity amplitudes are reasonably well known only 
for the photoproduction of the  prominent nucleon  resonances. 
In the case of electro-production accurate data are rare 
(for a review see e.g. \cite{Stoler}). 
We restrict ourselves to the dominant low mass 
resonances $\Delta(1232)$, $S_{11}(1535)$, and $D_{13}(1520)$. 
Our parametrizations of the corresponding helicity amplitudes 
are summarized in 
eqs.(\ref{eq:helicity_ampl_1},\ref{eq:helicity_ampl_2}), 
with parameters given in Table 1.

At low center of mass energies the excitation of the 
$\Delta(1232)$ resonance is of particular importance. 
At small $Q^2$ it is dominated by a magnetic dipole 
transition which implies $A_{3/2}/A_{1/2} \approx \sqrt{3}$.
Indeed, for real photons one finds $A_{3/2}/(\sqrt{3} A_{1/2})\approx 1.064 \,$
\cite{PDG}. 
At large momentum transfers, $Q^2 \gg 1$ GeV$^2$,  perturbative QCD gives  
$A_{3/2}/A_{1/2} \sim 1/Q^2$. 
However, it has been observed that even at $Q^2 = 3$ GeV$^2$ the  
magnetic dipole transition still dominates by far \cite{Burkert:1995hx}. 
We can therefore assume  $A_{3/2}/A_{1/2} \approx  const.$ for 
$Q^2 \,\lsim  \,3$ GeV$^2$. 
The $Q^2$-dependence of $A_{1/2}$ and $A_{3/2}$ 
is  then extracted from an analysis of the  $Q^2$-dependence 
of the transverse amplitude $|A_{T}|$ \cite{Stoler}. 

The $S_{11}$ resonance has spin $1/2$, so that 
the helicity amplitude  $A_{3/2}$ is absent. 
We constrain the parametrization of the amplitude $A_{1/2}$ 
by the photo- and electro-production data from ref.\cite{Breuker:1982nw}. 

For the $D_{13}(1520)$ the amplitude  $A_{1/2}$  is found 
to be very small at $Q^2 = 0$.  Here $A_{3/2}$ dominates. 
On the other hand  data require $A_{1/2} > A_{3/2}$
for $Q^2 > 1$ GeV$^2$\cite{Stoler}.

The parametrization in 
eqs.(\ref{eq:helicity_ampl_1},\ref{eq:helicity_ampl_2})
agrees with the present, albeit quite rough, 
empirical  information on the $Q^2$-dependence of the 
asymmetry ${\cal A}_1$ and the individual helicity amplitudes \cite{Stoler}: 
%
\begin{equation} \label{eq:helicity_ampl_1}
|A_T|  = \left(|A_{1/2}|^2+|A_{3/2}|^2\right)^{1/2}  
= \,C  \, \exp[- B\, Q^2], 
\end{equation}
and 
\begin{equation} \label{eq:helicity_ampl_2}
|A_{1/2,3/2}|= \sqrt{\frac{1\pm {\cal A}_1}{2}}\,|A_T|,\,\, 
\rm{with}\;\;\;\; {\cal A}_1 = \frac{|A_{1/2}|^2 - |A_{3/2}|^2}
{|A_{1/2}|^2 + |A_{3/2}|^2},
\end{equation}
with parameters given in table 1.

\begin{table}[h]\label{tabelle1}
\begin{center}
\begin{tabular}{|l|l|l|l|l|}\hline
 &$C/\rm{GeV}^{-1/2}$&$B/\rm{GeV}^{-2}$&${\cal A}_1$&$\Gamma/\rm{GeV}$
\\ \hline\hline
$\Delta(1232)$&$0.293$&$0.6$&$-0.545$&$0.12$\\ \hline
$S_{11}(1535)$&$0.07$&$0.17$&$1.0$&$0.15$\\ \hline
$D_{13}(1520)$&$0.16$&$0.83$&$1-\exp[0.3-1.5\; Q^2/\rm{GeV}^2]$&$0.12$\\ \hline
\end{tabular}
\vspace*{0.5cm} 
\caption{
Parameters for the helicity amplitudes in 
eqs.(\ref{eq:helicity_ampl_1},\ref{eq:helicity_ampl_2}).
The resonance widths $\Gamma$, which enter in eq.(\ref{eq:photon-nucleon_cross_sections}), 
are taken from  ref.\protect\cite{PDG}.}
\end{center}
\end{table}

The interference term $\sigma^{LT}$ involving the longitudinal and 
transverse photon-nucleon amplitudes 
is fairly unknown. 
Nevertheless, unpolarized scattering constrains the 
asymmetry ratio:
\begin{equation}
{\cal A}_2 = \frac{2\sigma^{LT}_{1/2}}{\sigma^T_{1/2}+\sigma^T_{3/2}} 
< \sqrt {R(x,Q^2)},   
\end{equation}
with $R =2\sigma_{1/2}^L/ (\sigma_{1/2}^T + \sigma_{3/2}^T)$. 
In the resonance region one finds on average $R = 0.06 \pm 0.02$ 
for $1 \,\rm{GeV}^2 <  Q^2 < 8 \,\rm{GeV}^2$ and 
$W < 1.7$ GeV \cite{keppel}. 
Some fraction of this  value is due to incoherent background 
contributions and not related to the excitation of single nucleon resonances. 
In the following we use ${\cal A}_2 = 0.08$. 
As a matter of fact, at $Q^2 > 1$ GeV$^2$,  ${\cal A}_2$ contributes only very
little to the structure function moments to be discussed later: 
changing  ${\cal A}_2$ by $100\%$ modifies 
our results for $g_1^{(1)}$ by less than $2\%$.

At low energies, $W < 1.7$ GeV,  
the structure function $g_1$ receives  contributions also 
from non-resonant (multi-)meson  production. 
However, hardly any empirical information is available here.
We use a linear interpolation in the squared photon-nucleon 
center of mass energy $W^2$ which 
connects the inelastic threshold $W^2= (M+m_{\pi})^2$ with 
experimental data at $W > 1.7$ GeV.

Having modeled the 
structure function $g_1^p$ at small  center-of-mass energies, we continue to 
$W > 1.7$ GeV, where we use a  
parametrization from  ref.\cite{E143} which reproduces data 
in the deep-inelastic region. 
This, finally, completes our model for $g_1^p$. 

\begin{figure}
\centering{\ \epsfig{figure=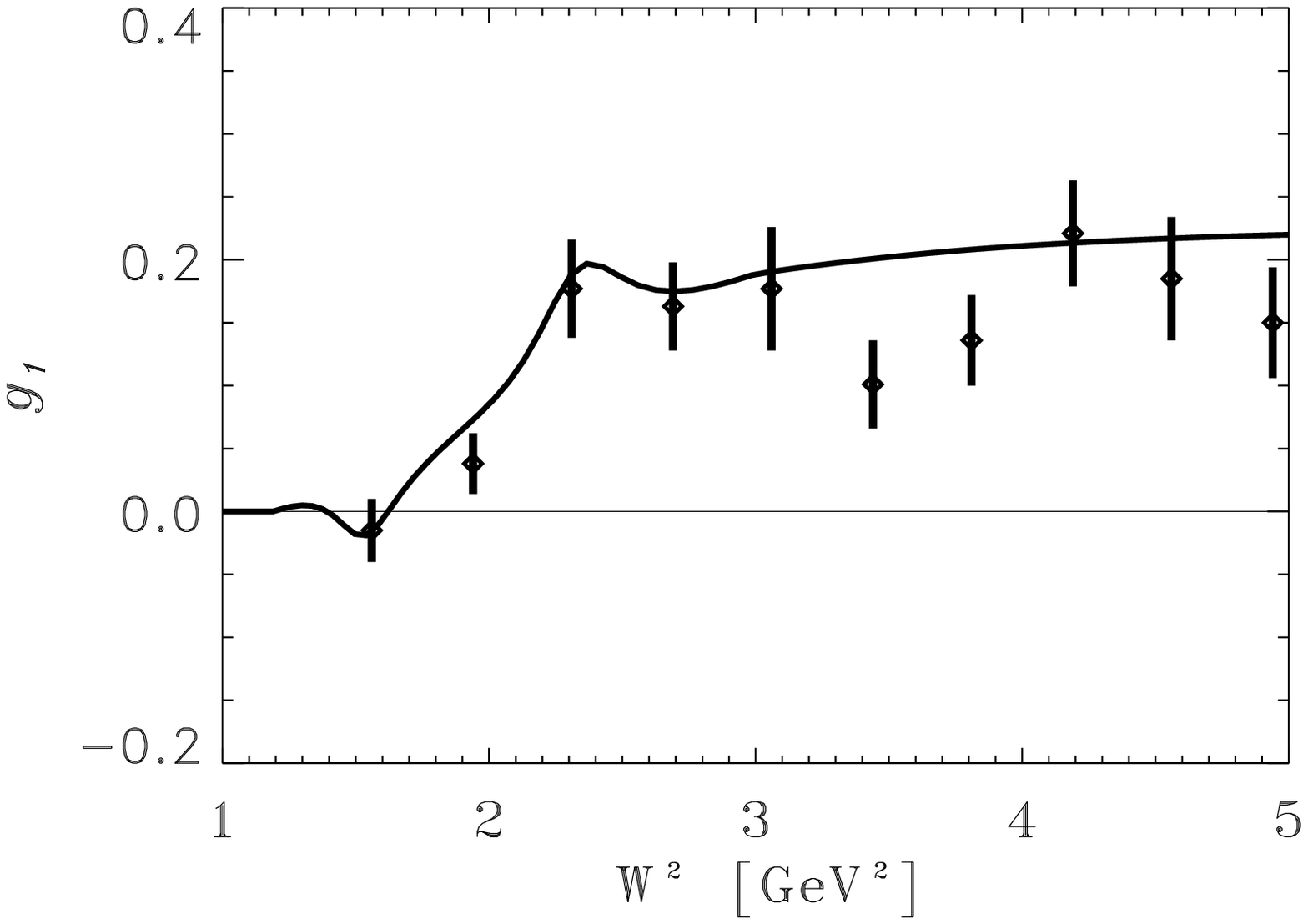,height=10cm}}
\caption{\label{Fig0}
The proton spin structure function $g_1^p$ at $Q^2\;=\;1.2\;{\rm GeV}^2$ as calculated 
from our model in section \ref{sed:Model}. The data 
are taken from  ref.\cite{E143}. 
} 
\end{figure}

\begin{figure}
\centering{\ \epsfig{figure=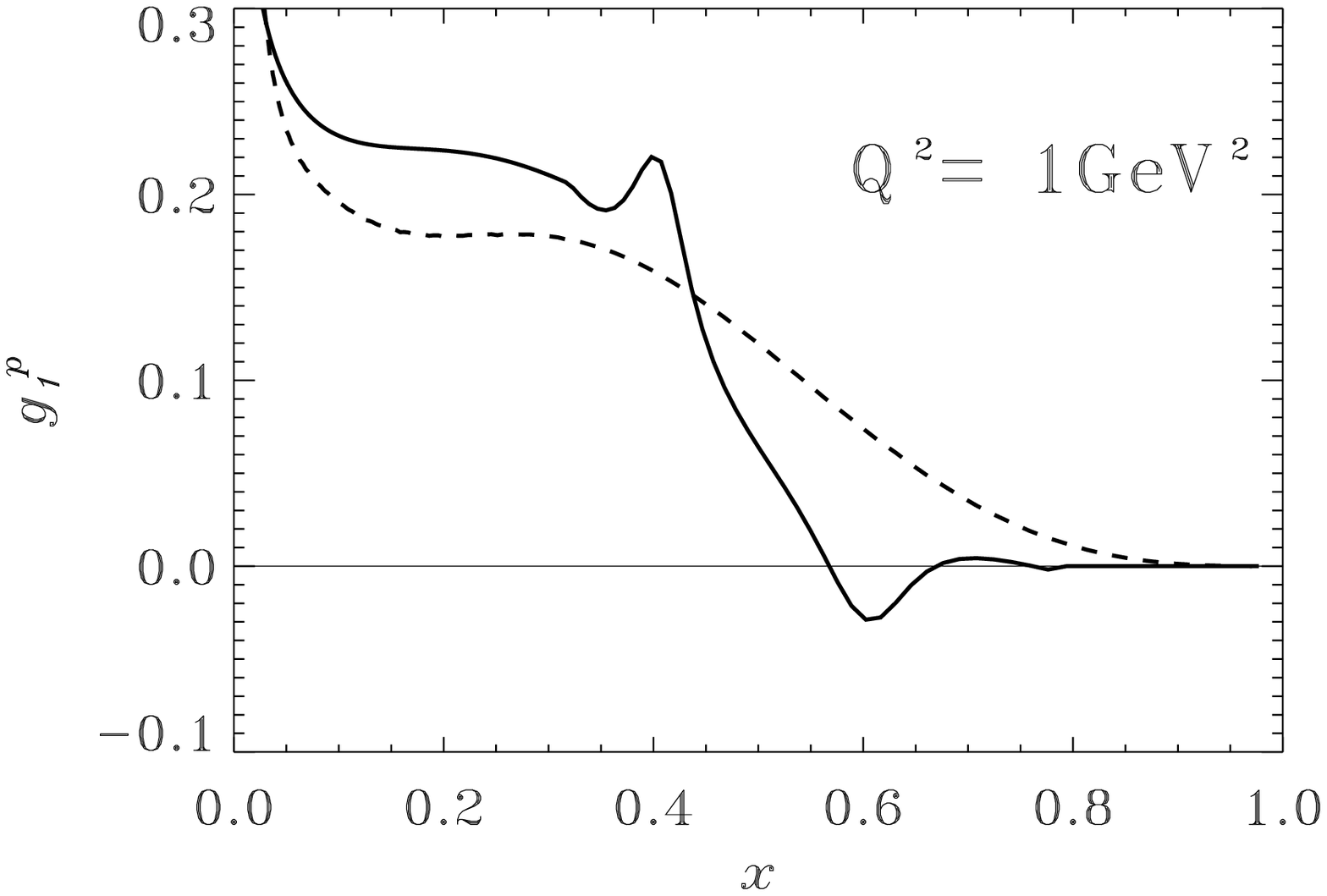,height=5cm}}
\centering{\ \epsfig{figure=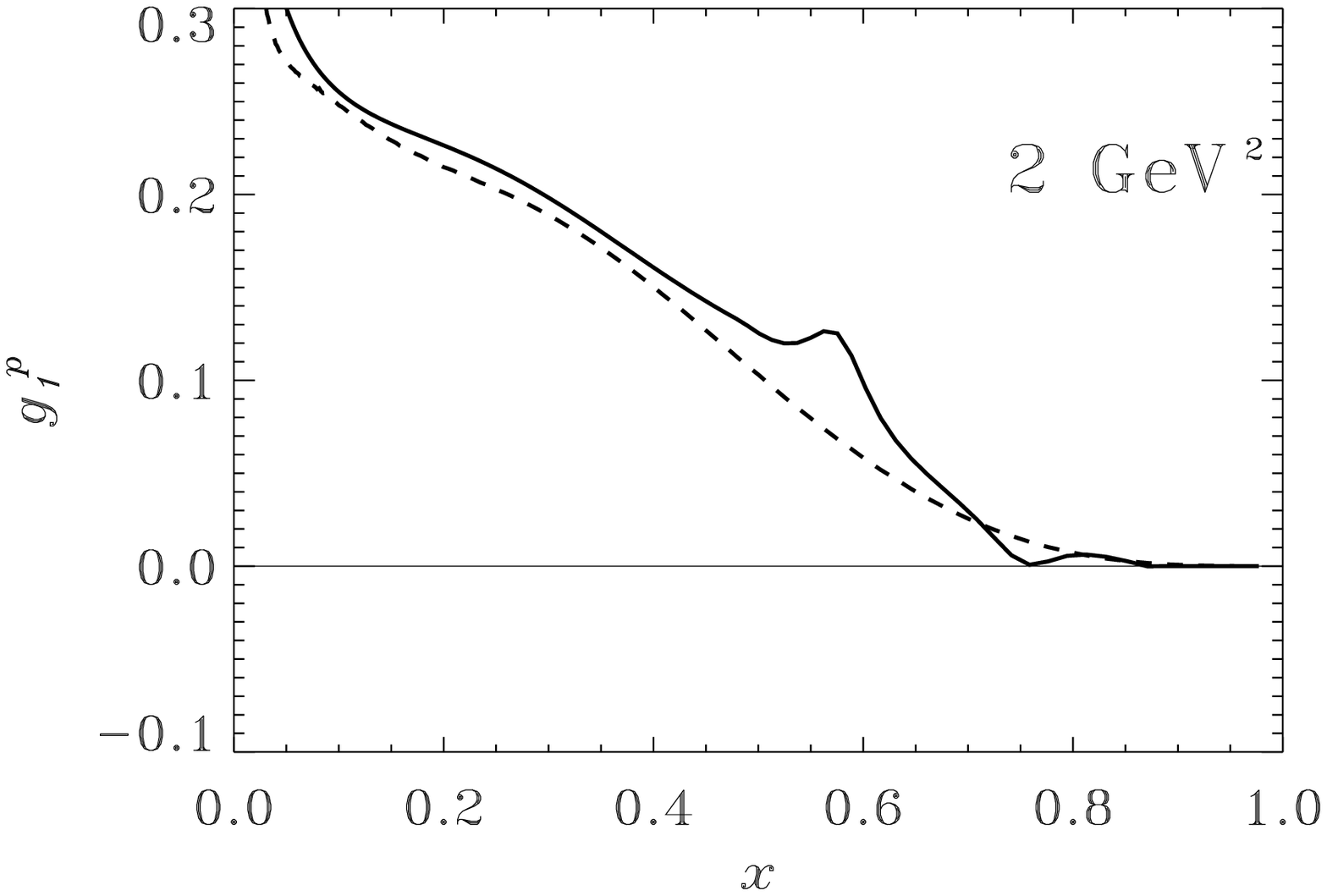,height=5cm}}
\centering{\ \epsfig{figure=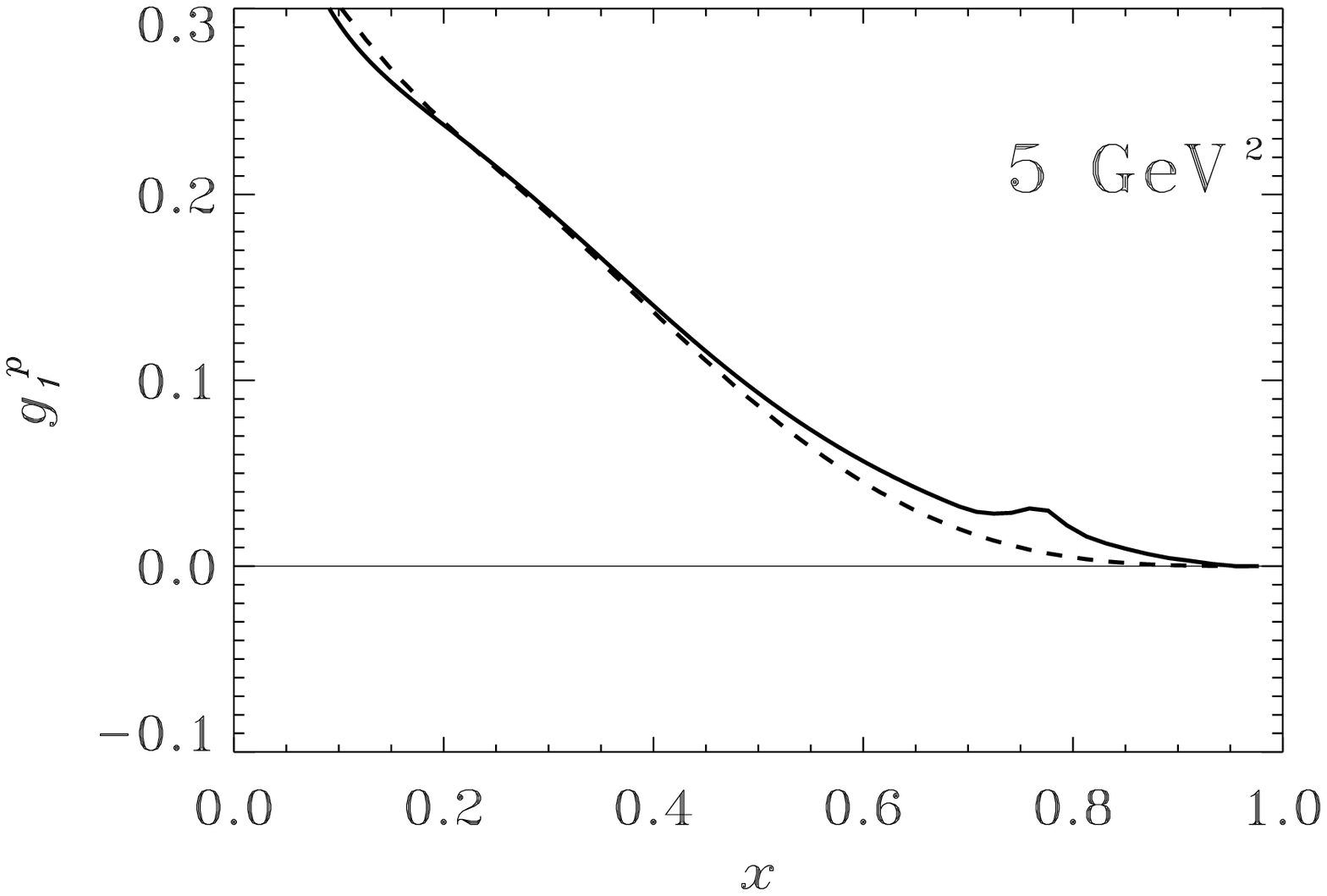,height=5cm}}
\centering{\ \epsfig{figure=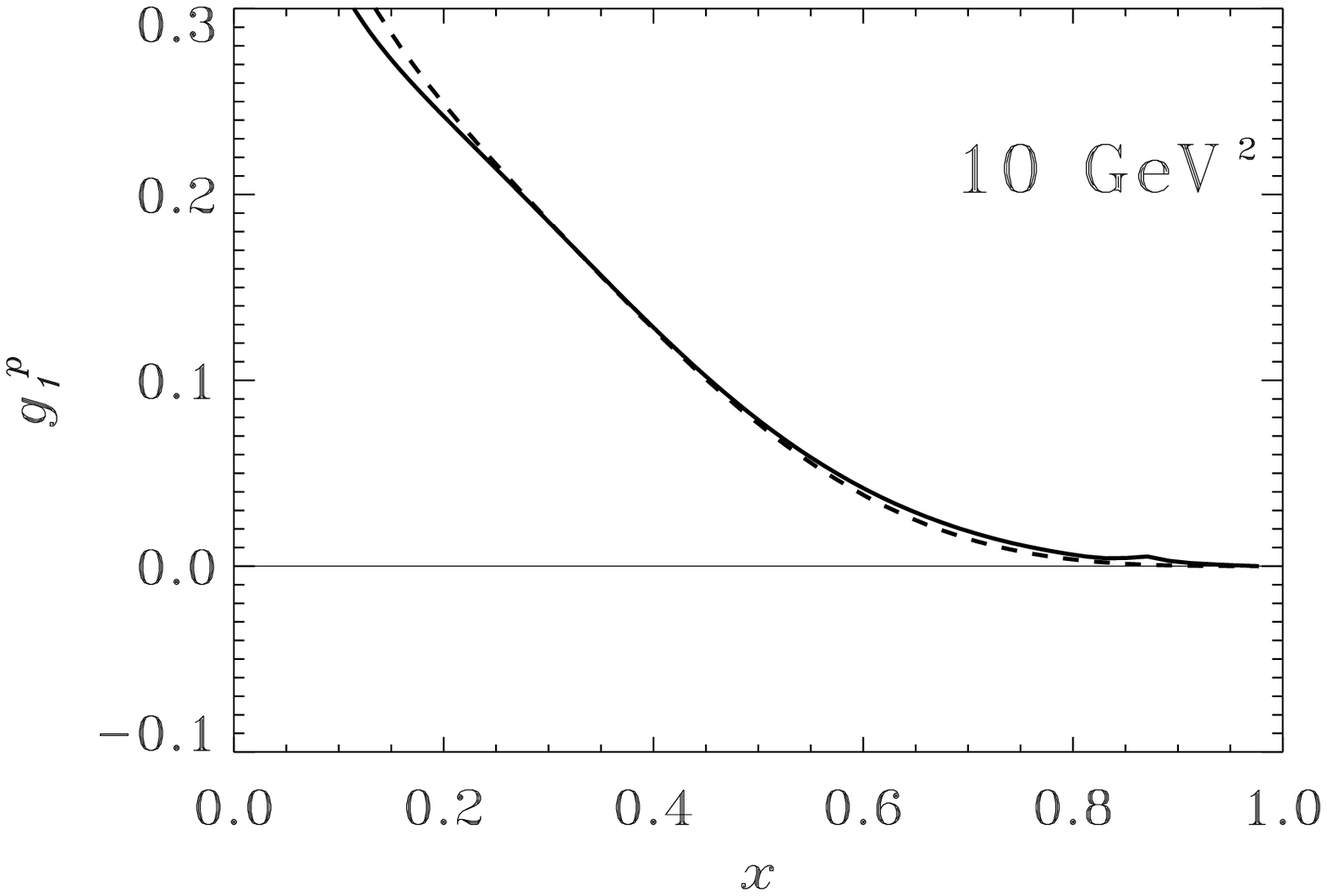,height=5cm}}
\caption{\label{Fig1}
The $x$-dependence of $g_1^p$ for $Q^2=1,2,5$ and $10 \;{\rm GeV}^2$. 
The full lines show results of the model developed in section \ref{sed:Model}, the dashed lines correspond 
to the twist-$2$ parametrization of  ref.\cite{Gehrmann}
}
\end{figure}

In Fig.\ref{Fig0} we compare our model 
with recent $g_1$  data from the E143 collaboration 
taken at  $Q^2 = 1.2$ GeV$^2$.  
Within the  admittedly large experimental errors 
good agreement is found.
A comparison of  $g_1^p$,  
calculated with our model,  
and a  parametrization of its deep-inelastic twist-$2$ 
part \cite{Gehrmann} is shown in Fig.\ref{Fig1}. 
At  $Q^2\,\lsim \,2$ GeV$^2$,
significant deviations  are apparent.
The contributions of  the $S_{11}$ and $D_{13}$ resonances  
are located around $x\sim 0.4$ at $Q^2 = 1$ GeV$^2$, 
while the excitation of the $\Delta$ occurs at  $x\sim 0.6$. 
 As $Q^2$ increases the low mass nucleon resonance 
excitations become less important. 
Furthermore, the contribution of nucleon 
resonances moves  towards larger values of $x$, as one can see 
from the fact that the 
squared invariant mass of a particular nucleon excitation is fixed at  
$W^2 = M^2 + Q^2 (1-x)/x$.  
Finally, at $Q^2=10$ GeV$^2$ our model 
coincides with the leading twist parametrization of  ref.\cite{Gehrmann}.

\section{Analysis of moments of structure function}

In this section we discuss  the first moments of the proton structure 
function $g_1$ as obtained from the model previously described. 
In particular, we investigate the importance of contributions 
from elastic scattering, resonance production, 
target mass corrections, and true higher twist. 

The elastic contribution, corresponding to the kinematic 
limit $x=1$, is determined by the 
Pauli and Dirac electromagnetic form factors of the nucleon as follows:
\begin{equation}\label{eq:g1_el}
\left.g_1^{(n)}(Q^2)\right|_{el} = 
\frac{1}{2}F_1(Q^2)\left[F_1(Q^2)+F_2(Q^2)\right].
\end{equation}
In our numerical analysis we use parametrizations from  
ref.\cite{MeissnerFormf}.

\subsection{Resonance contributions}
\label{ssec:res}

\begin{figure}
\centering{\ \epsfig{figure=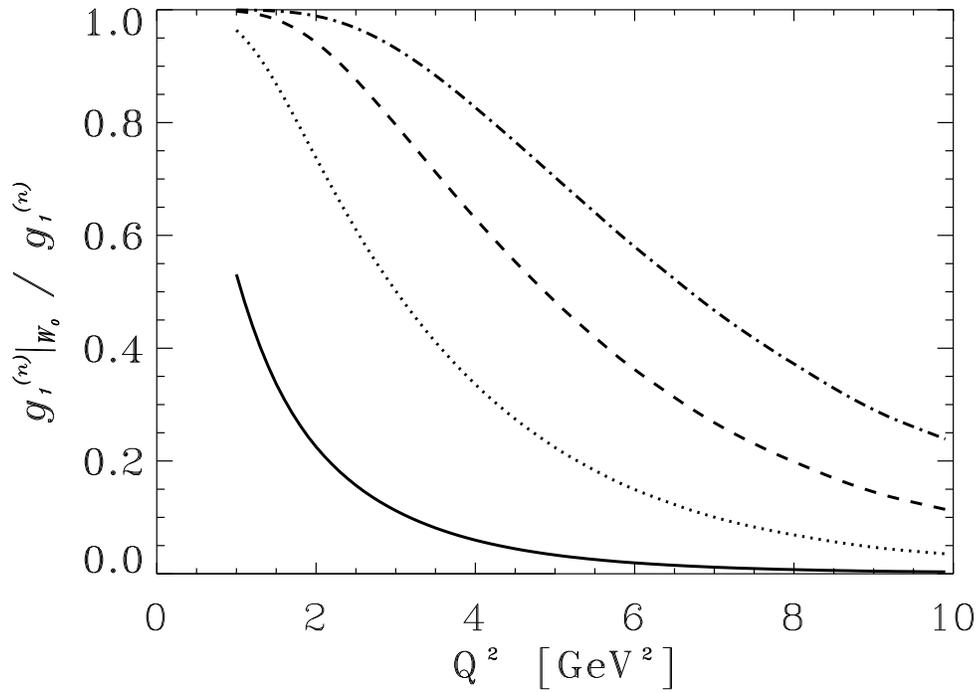,height=10cm}}
\caption{\label{Fig2}
Contribution to the moments $g_1^{(n)}$ from low mass excitations with $W < W_0 = 2$ GeV, normalized 
to the full moments. Shown are the $1^{\rm{st}}$ (full), 
$3^{\rm{rd}}$ (dotted), $5^{\rm{th}}$ (dashed), and 
$7^{\rm{th}}$ (dash-dotted) moments. 
} 
\end{figure}

In order to investigate the role of low-mass nucleon excitations 
it is useful to introduce the ratio 
\begin{eqnarray}
\frac{\left.g_1^{(n)}(Q^2)\right|_{W_0}}{g_1^{(n)}(Q^2)}
=\frac{\int_{x_0}^1{\rm d}xx^{n-1}g_1(Q^2)}{
\int_{0}^1{\rm d}xx^{n-1}g_1(Q^2)},\quad\quad\
{\rm with}\;\;x_0=x(W=W_0). 
\end{eqnarray}
In the numerator we sum over all contributions of nucleon resonances and  
non-resonant multi-meson excitations with invariant mass 
$M_X < W_0  = 2$ GeV. In addition we always include the elastic part 
(\ref{eq:g1_el}). 
Figure \ref{Fig2}  shows that these low mass contributions to 
$g_1^{(n)}$ are quite sizable, especially for higher moments. 
For example, at $Q^2 = 2$ GeV$^2$ they are responsible for 
about $20\%$ of the first moment $g_1^{(1)}$, while they 
account for $75\%$ of $g_1^{(3)}$.
With increasing $n$ the role of low-mass excitations becomes 
evidently more pronounced. 
At the same time, the influence of the low-mass part of the spectrum also
increases with decreasing $Q^2$. 
Roughly speaking, for $Q^2 < 2 n\;{\mbox{GeV}}^2$ low-mass excitations with
$W\;<\;2\;{\rm GeV}$ account for 
more than $10\%$ of $g_1^{(n)}$. 
At large  $Q^2$ continuum contributions with $W > W_0$ 
take over. 
A similar observation has been made in an analysis of unpolarized 
lepton scattering \cite{JiUnrauUnpol}.

\subsection{Higher twist analysis}
 
In order to extract the genuine higher twist coefficients $f_n$ 
from the structure function moments $g_1^{(n)}$  
one has to subtract twist-$2$ contributions and 
target mass corrections from each given moment. Returning to 
eq.(\ref{eq:g1n_HT}) we have:
\begin{eqnarray}\label{extract}
g_1^{(n)}(Q^2)|_{\rm ht} &=& g_1^{(n)}(Q^2)- \frac{1}{2}a_{n-1}(Q^2)-
\frac{M^2}{Q^2}\frac{n(n+1)}{2(n+2)^2}
\left(n\, a_{n+1}(Q^2)+4 \,d_{n+1}(Q^2)\right)\\  
&=&f_{n+1} \frac{4}{9}\frac{M^2}{Q^2}+{\cal O} \left( \frac{M^4}{Q^4}\right). 
\end{eqnarray}
In the following we consider the first three moments, $n=1,3,5$. 
We compare results obtained  from our  model for $g_1^p$ 
with the twist-$2$ contributions $a_{n-1}/2$ from the 
NLO parametrization of  ref.\cite{Gehrmann}. 
We also study the influence of target mass effects. 
Finally we discuss different contributions to the 
higher twist part $g_1^{(n)}(Q^2)|_{\rm ht}$.

\begin{figure}
\centering{\ \epsfig{figure=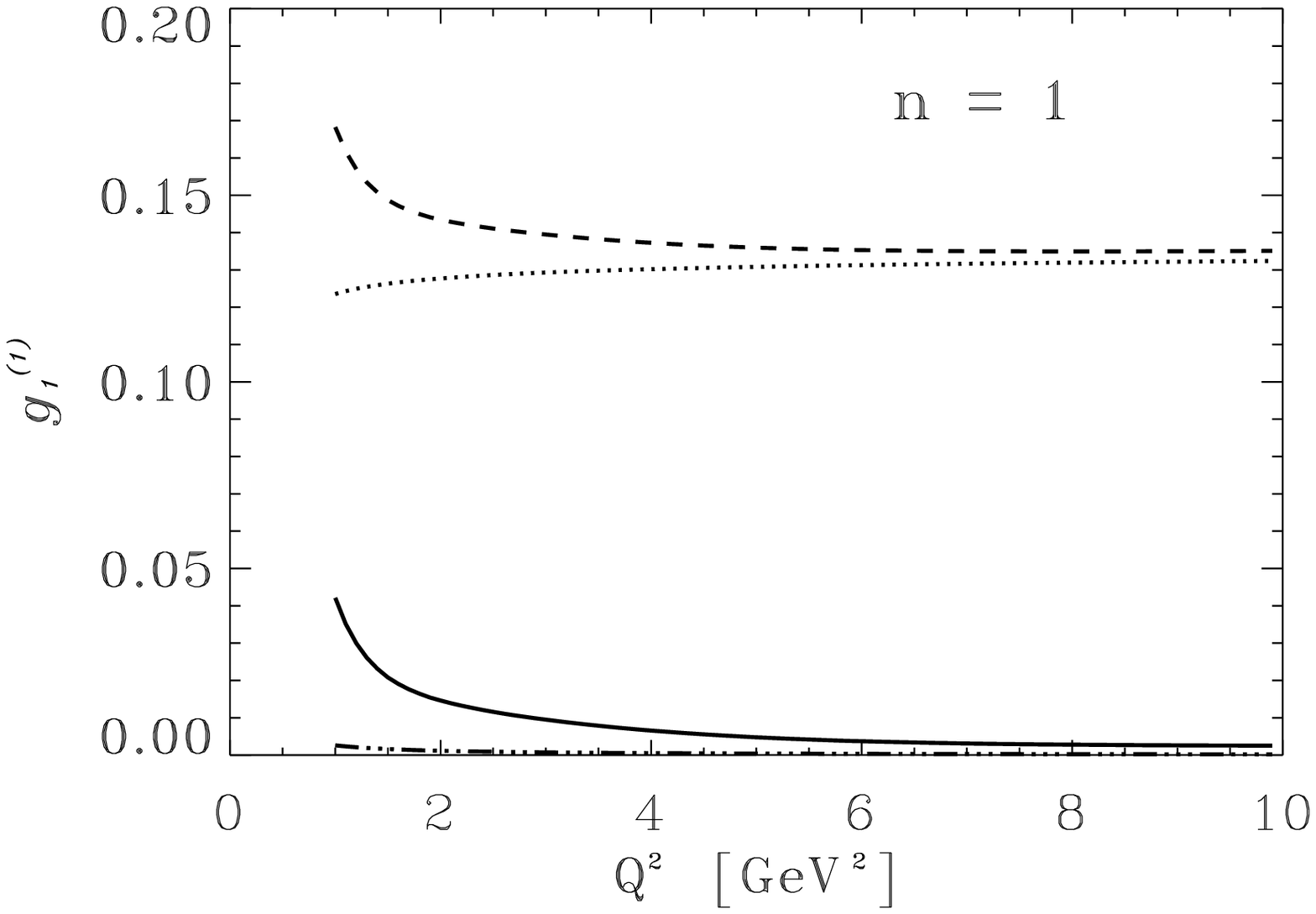,height=7cm}}
\centering{\ \epsfig{figure=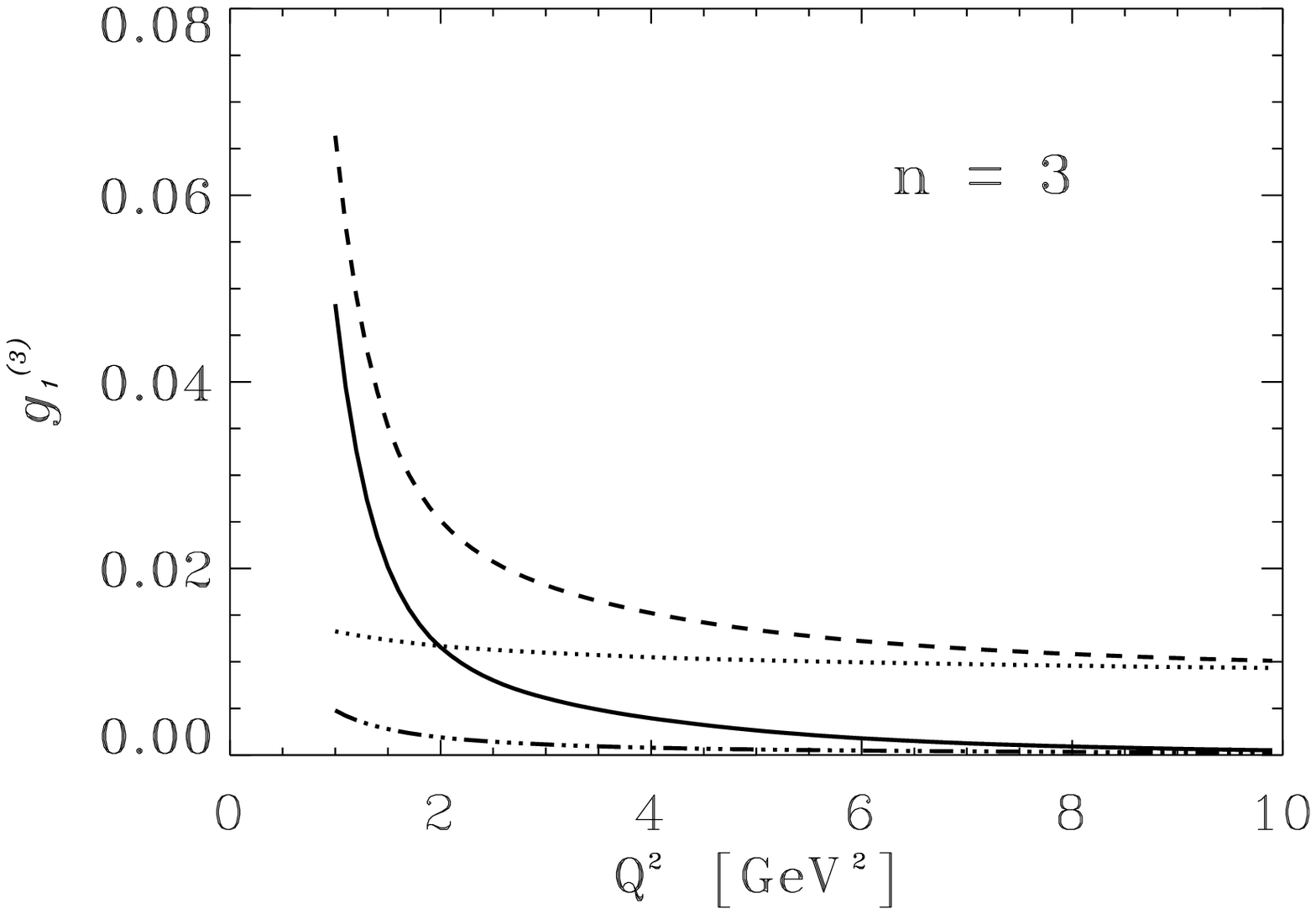,height=7cm}}
\centering{\ \epsfig{figure=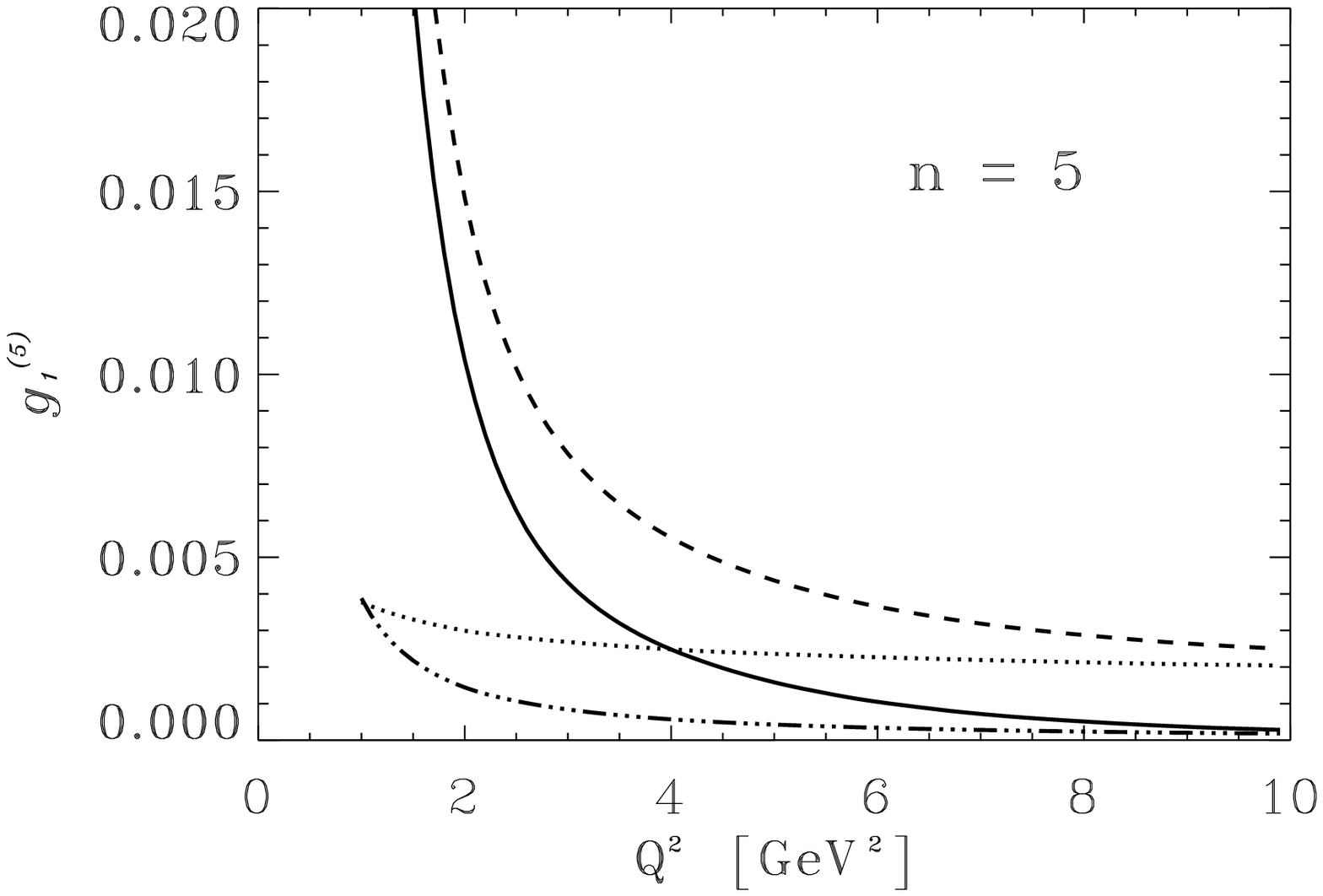,height=7cm}}
\caption{\label{Fig3}
The moments $g_1^{(n)}$ and their decomposition (\ref{extract}) 
for $n=1,3$ and $5$. 
Shown are: $g_1^{(n)}$ (dashed), the twist-2 part 
$a_{n-1}/2$  (dotted), the target mass corrections (dash-dotted), and
the higher twist piece $g_1^{(n)}|_{\rm ht}$ of Eq.(\ref{extract})(full).
}
\end{figure}

In Fig.\ref{Fig3} we compare the full moments $g_1^{(n)}$ with the 
leading twist parts, $a_{n-1}/2$, and the higher twist components 
$g_1^{(n)}(Q^2)|_{\rm ht}$. 
At small  $Q^2$ one observes 
significant differences between  $g_1^{(n)}$  and $a_{n-1}/2$.
In particular  one finds 
$g_1^{(n)}(Q^2)|_{\rm ht} > 0.1 \, a_{n-1}/2$  
for $Q^2 < 2,\,4,\,10$ GeV$^2$ and $n=1,\,2,\,3$, respectively.  
The region where higher twist becomes  important 
depends obviously on the moment $n$.
For fixed $Q^2$ the difference between  $g_1^{(n)}(Q^2)$ 
and $a_{n-1}/2$ increases with $n$. 
This is easily understood since contributions of low-mass 
nucleon excitations are enhanced in higher moments as pointed 
out in the previous section. 
Also shown in Fig.\ref{Fig3} is the size of target mass effects. Since the 
coefficients $d_n$ are not known accurately
we use $d_n = 0$ which is compatible with present data \cite{E143,E154}  
and corresponds to the Wandzura-Wilczek conjecture \cite{Wandzura}. 
For this choice target mass effects are indeed small. 
As an example, at $Q^2=2$ GeV$^2$ and n=1,3,5 they amount to less than $10\%$ of the 
higher twist part. 
To estimate the uncertainty of this result we  also use 
$d_n$ obtained from eq.(\ref{eq:dn_g2}) for  $g_2(x) = 0$.
In this case target mass effects increase significantly and lead to  
a decrease of $g_1^{(n)}(Q^2)|_{\rm ht}$ by about $30\%$. 
High precision data on the spin structure function $g_2$, which 
are of course interesting in their own right, are therefore 
an important ingredient in the QCD analysis of $g_1$ itself. 

\begin{figure}
\centering{\ \epsfig{figure=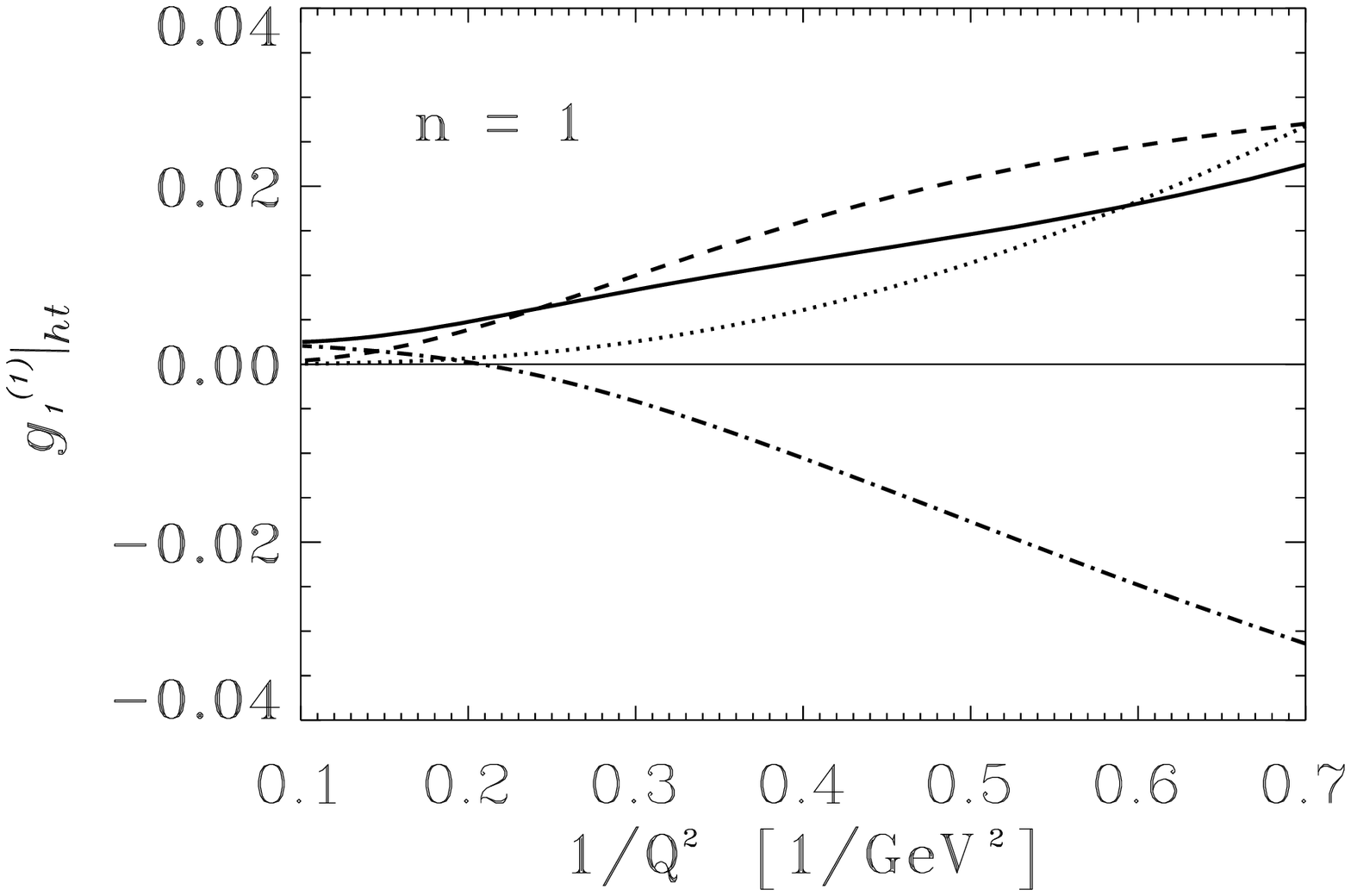,height=6.5cm}}
\centering{\ \epsfig{figure=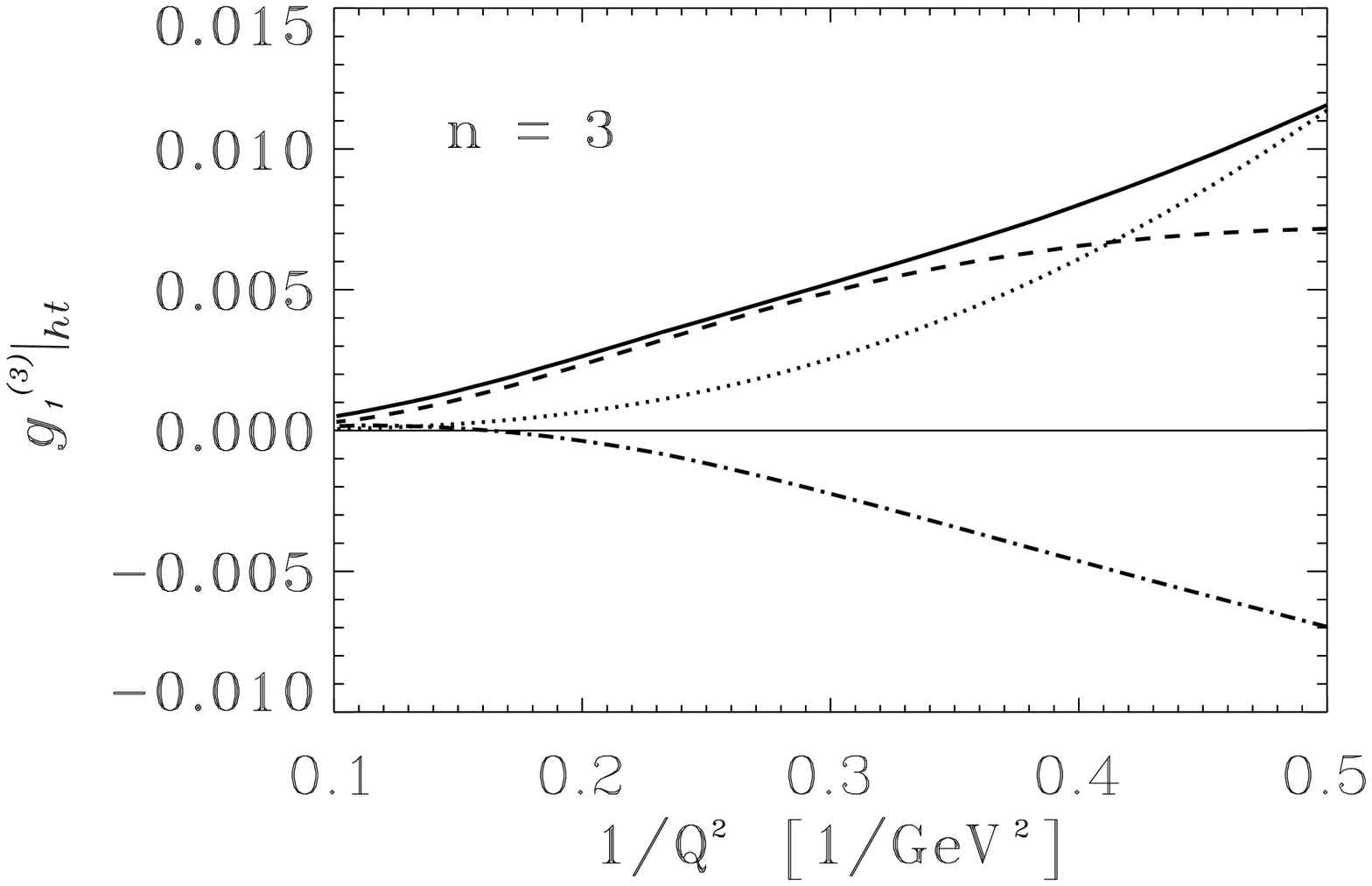,height=6.5cm}}
\centering{\ \epsfig{figure=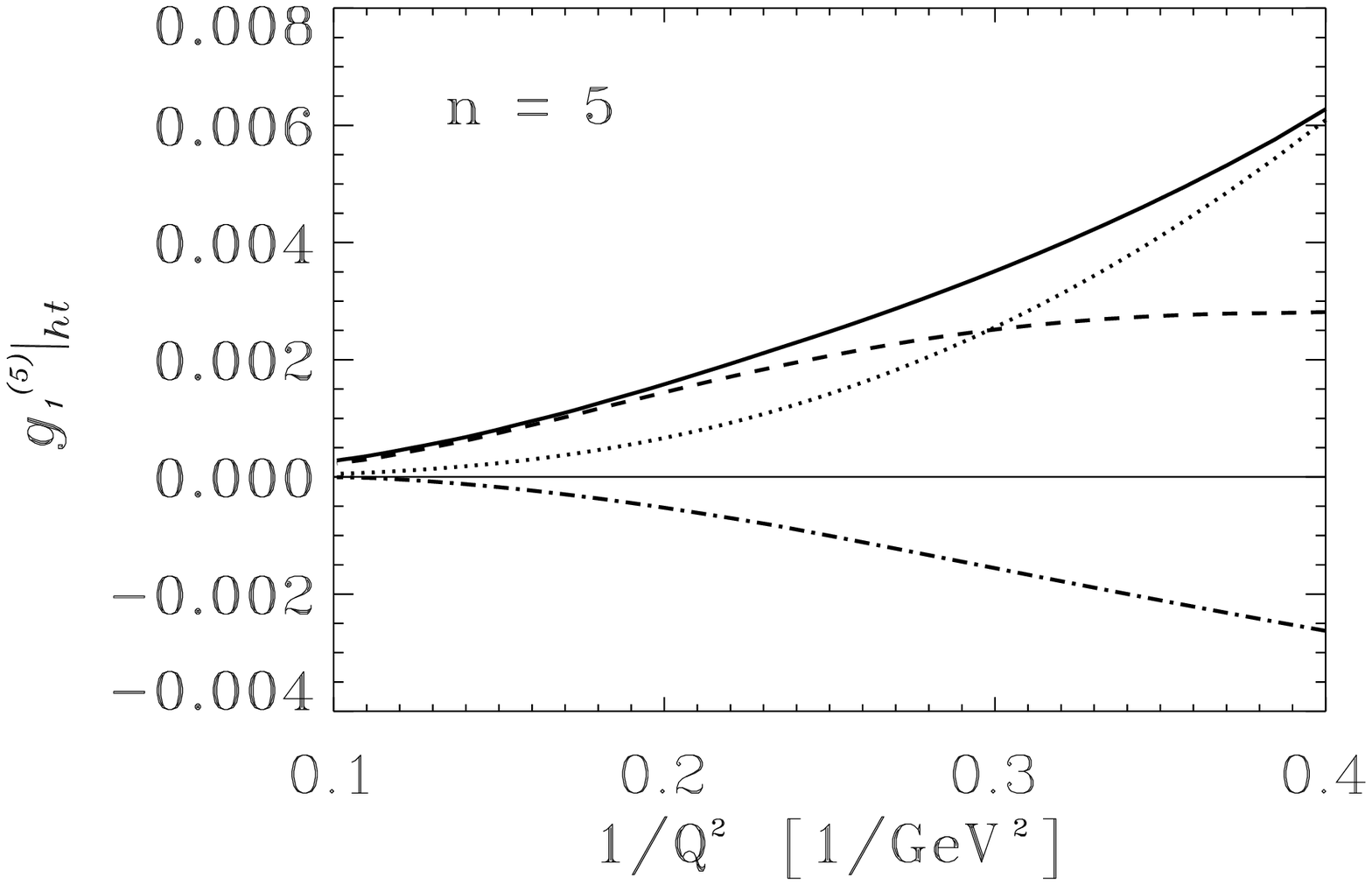,height=6.5cm}}
\caption{\label{Fig4}
Higher twist components $g_1^{(n)}|_{\rm ht}$ 
for $n=1,3$ and $5$ and their decomposition, plotted as a function of $1/Q^2$. 
Shown are $g_1^{(n)}|_{\rm ht}$ (full), 
the elastic part (dotted),   
contributions from low-mass excitations with $W<W_0 = 2$ GeV 
(dashed), and  the difference between the high-mass continuum with
$W>W_0$ and the twist-2 part (dash-dotted).
}
\end{figure}

Twist-4 contributions to $g_1^{(n)}$ are proportional to $1/Q^2$ (up to
logarithmic corrections). In order to have a closer look at these terms
it is instructive to plot the higher-twist moments $g_1^{(n)}(Q^2)|_{\rm ht}$
versus $1/Q^2$, as done in Fig. \ref{Fig4}.
Their approximately linear behavior indicates that twist-4 contributions
play indeed a dominant role in $g_1^{(n)}(Q^2)|_{\rm ht}$. Neglecting terms of twist-6
and higher gives $f_2^p\simeq 0.1$ at $Q^2\;=\;2\;{\rm GeV}^2$, which agrees
quite well with the analysis of  ref.\cite{JiMel}.
Further estimates can be found in ref.\cite{Ji:1995bs}.

In the same figure we show the separate contributions to $g_1^{(n)}(Q^2)|_{\rm
ht}$ from elastic scattering and from low-mass excitations with 
$(M+m_{\pi})< W <2\;{\rm GeV}$. Evidently, none of these contributions is
small, in fact they all
are of the same order of magnitude as $g_1^{(n)}(Q^2)|_{\rm
ht}$ itself.

These observations emphasize the need for high-precision measurements
especially in the resonance region. Upcoming data from TJNAF \cite{JEFPROP} are
certainly welcome here. Figure \ref{Fig4} also points to the crucial role
played by the elastic piece (\ref{eq:g1_el}). Its proper treatment requires
accurate information on the nucleon electromagnetic form factors in the range
$1.5\;{\rm GeV}^2<Q^2<10\;{\rm GeV}^2$.

For the higher moments with $n=3,5$ the kinematic window in which twist-4
contributions dominate, that is, where 
$g_1^{(n)}(Q^2)|_{\rm ht}$ behaves linearly with $1/Q^2$, moves successively to
higher $Q^2$. Again the contributions from elastic, resonant and non-resonant
scattering all turn out to be of similar importance.




\subsection{Parton-hadron duality}

With decreasing $Q^2$ the higher twist contributions eventually 
reach the magnitude of the leading twist parts. 
As a consequence the twist expansion (\ref{eq:twist_expansion}) breaks down. 
Our model for $g_1^p$ can be used to suggest
where this transition  takes place: for a given moment $g_1^{(n)}$ 
higher twist contributions amount to less than  $50\%$ of 
the leading twist ones if $Q^2  > n $ GeV$^2$. 
On the other hand, we have learned in section \ref{ssec:res} that 
low mass excitations account for more than $10\%$ of $g_1^{(n)}$ 
if $Q^2 < 2 n\;{\mbox{GeV}}^2$.

This indicates a region of $n$ and $Q^2$ in which perturbative higher twist
corrections coexist with resonance contributions. The resonance terms are
significant, and the transition amplitudes involving these resonances introduce
powers of $1/Q^2$ in just such a way that they follow the deep-inelastic, large
$Q^2$ behaviour of $g_1^p$.


Such a behavior is known as parton-hadron duality, a notion introduced  
by Bloom and Gilman for the unpolarized structure function $F_2$. 
A QCD explanation of this phenomenon has first been offered  in  
 ref.\cite{Georgi:1976} and was further elaborated  in  ref.\cite{DeRujula77}. 
According to our results similar arguments apply to 
polarized lepton-nucleon scattering.

\section{Summary}

\begin{itemize}
\item[i)]
Contributions from the region of the nucleon resonances are an essential
ingredient in the "higher-twist" analysis of the spin structure function $g_1$.
Their effects are clearly visible in $g_1^p$ even at $Q^2$ as large as $5\;{\rm
GeV^2}$. For example, low mass excitations with $W\;<\;2\;{\rm
GeV}$ account for more than $50\%$ of the 3rd moment and more than $80\%$ of
the 5th moment of $g_1^p$ in the range $Q^2\;\lsim\; 3\;{\rm
GeV^2}$.

\item[ii)]
We have pointed to the importance of the elastic scattering ($x=1$) part in a
consistent moment analysis of $g_1$. Without inclusion of this elastic part, an
extraction of higher-twist terms would not be meaningful.

\item[iii)]
We observe a coexistence of resonance contributions and perturbative
higher-twist corrections in a window, roughly framed by $n\;{\rm
GeV}^2\;<\;Q^2\;<\;2n\;{\rm
GeV}^2$, where $n=1,3,5,\dots$ denotes the moment of $g_1$. The understanding
of this coexistence region in terms of parton-hadron duality is an interesting
issue. Precision data from TJNAF will help clarifying these questions in the
near future.

\end{itemize}

{\bf Acknowledgments}:

The authors wish to acknowledge helpful discussions with K.A. Griffioen and 
L. Mankiewicz.


\newpage

\pagebreak

\pagebreak

\pagebreak

\pagebreak

\pagebreak

\end{document}